\documentclass[sigconf]{acmart}
\renewcommand\footnotetextcopyrightpermission[1]{} 
\def\BibTeX{{\rm B\kern-.05em{\sc i\kern-.025em b}\kern-.08em
    T\kern-.1667em\lower.7ex\hbox{E}\kern-.125emX}}

\usepackage{booktabs} 
\usepackage{xurl} 
\usepackage{amsfonts}
\usepackage{tabularx}
\usepackage{siunitx}
\usepackage{enumitem}
\usepackage{stfloats}
\usepackage{geometry}
\usepackage{multirow}
\usepackage{makecell}
\usepackage{tcolorbox}
\usepackage{stackengine,graphicx}
\usepackage{xspace}
\usepackage{caption}
\usepackage{subcaption}
\usepackage{graphicx}
\usepackage{makecell}
\usepackage{multirow}
\usepackage{pgf}
\usepackage{url}
\usepackage{color}
\newcounter{querycount}
\usepackage{enumerate}

\makeatletter
\newcommand{\myitem}[1]{%
\item[#1]\protected@edef\@currentlabel{#1}%
}
\makeatother

\usepackage{verbatimbox}

\setcopyright{none}

\makeatletter
\renewcommand\@formatdoi[1]{\ignorespaces}
\makeatother

\begin{document}

\title{Modelling Legislative Systems into Property Graphs to Enable Advanced Pattern Detection}

\author{Andrea Colombo}
\affiliation{%
  \institution{Politecnico di Milano}
  \city{Milan}
  \country{Italy}
}
\email{andrea1.colombo@polimi.it}

\author{Anna Bernasconi}
\affiliation{%
  \institution{Politecnico di Milano}
  \city{Milan}
  \country{Italy}
}
\email{anna.bernasconi@polimi.it}

\author{Stefano Ceri}
\affiliation{%
  \institution{Politecnico di Milano}
  \city{Milan}
  \country{Italy}
}
\email{stefano.ceri@polimi.it}


\begin{abstract}
Legislative systems face growing complexity due to the ever-increasing number of laws and intricate interdependencies between them. Traditional methods of storing and analyzing legal systems, mainly based on RDF, struggle with this complexity, hindering efficient knowledge discovery, as required by domain experts.
In this paper, we propose to model legislation into a property graph, where edges represent citations, modifications, and abrogations between laws and their articles or attachments, both represented as nodes and edges with properties. 
As a practical use case, we implement the model in the Italian legislative system. First, we describe our approach to extracting knowledge from legal texts. To this aim, we leverage the recently internationally adopted XML law standard, Akoma Ntoso, to parse and identify entities, relationships and properties.
Next, we describe the model and the schema implemented using Neo4j, 
the market-leading graph database management system. The schema is designed to capture the structure and hierarchy of laws, together with their interdependencies. We show how such a property graph enables an efficient answer to complex and relevant queries previously impractical on raw text. 
By leveraging other implementations of the Akoma Ntoso standard and the proposed property graph approach, we are confident that this work will facilitate a comprehensive comparison of legislative systems and their complexities.
\end{abstract}




\maketitle

\vspace{-1mm}

\section{Introduction}

The adoption of emerging database and knowledge representation technologies for legislative systems presents a crucial step toward enhancing accessibility, comprehensibility, and efficiency within legal domains. The textual nature of laws presents a significant challenge when it comes to extracting or processing the content in a structured manner for automated analysis. 
Many efforts have been devoted to proposing appropriately expressive conceptual, machine-readable models of various aspects of general legal knowledge. 
Most of these proposals have been in the eXtensible Markup Language (XML) format, which is a semi-structured data model widely used in modeling textual data, thus capable of easily representing legislative texts and their content~\cite{lupo2007general}. Among them, we can recall the Legal Knowledge Interchange Format (LKIF)~\cite{hoekstra2007lkif} and LegalRuleML~\cite{legalRuleML}. 
More recently, we witnessed the official adoption by many international and national bodies of a common standard based on XML, Akoma Ntoso (AKN)~\cite{akomantoso}. It is an XML-based standard that aims to propose a generic model for acts, laws, or bills, such that it is flexible enough to represent, within the same schema, different legislative systems. Such a feature unlocks the possibility of creating generic pipelines that process distinct national legislations, with the goal of comparing the features of distinct countries.
For instance, by leveraging XML tags, the standard allows us to derive knowledge graphs, i.e., collection of documents or even parts of documents, namely articles, which may be linked in complex ways and used to derive statistics and metrics about the legislative systems~\cite{Sadeghian2018}.

In this context, the Semantic Web community has made crucial advancements by building ontologies as well as the RDF (Resource Description Framework) paradigm that can be used to represent legal information in RDF-based knowledge graphs~\cite{anelli2022navigating}. While such efforts are an important step toward linking knowledge bases by providing globally unique and resolvable identifiers over multiple domains, they are constrained to use the edge-labeled graph data model, whose RDF graphs are a special type of~\cite{sparqlcypher}.
RDF is based on triples, consisting of a subject, a predicate, and an object; triples can be considered statements describing the relationship between the subject and the object. 
While RDF graphs can be navigated through SPARQL~\cite{perez2009semantics}, (i) finding paths is not easily achievable, mainly because graph construction and path of edges cannot be naturally represented in the tabular result format of SPARQL~\cite{queryGraph,SPARQLResults} and (ii) their storage in the form of triples -- as independent artifacts -- precludes a rapid relationship traversal~\cite{robinson2015graph}. Additionally, each legislation tradition has some features that make it unique, and thus, a more flexible attribute representation than RDF single-value properties is preferable, as discussed in~\cite{Das2014ATO}. 
For instance, an edge connecting a source article with the target law might indicate a partial abrogation, thus having a type of reference, i.e., abrogation, but not denoting a complete scrap of the target law; an additional property on the edge can trivially represent such a feature.

The need for a computable and flexible representation of the knowledge about legislation that goes beyond the adoption of machine-readable standards or the Linked Data initiative has been inspired by our interactions with experts at the Einaudi Institute for Economics and Finance (EIEF~\cite{eief}), a research center based in Rome, Italy, that produces research on policy-relevant topics. 
Our requirements have been collected through a series of interviews, resulting in the need for a system that extracts non-trivial and aggregated information about the general trends in the production of legislation, which aims to understand the complexity of the legislative system.
We understood that these questions could only be answered by having a formal and interoperable representation of the data, as RDF-ones, but that also allows us to easily express more advanced graph and metric-related queries, accounting for system-specific features. To address this, we identified the most natural solution as a property graph database, which enables an easy derivation of metrics that evaluate the efficiency and impact of legislative production. By analyzing the structure and dynamics of the legislative property graph, researchers and policymakers can gain valuable insights into patterns of legislative activity, trends in lawmaking, and the effectiveness of legislative interventions. For example, we can get measures of legislative output (e.g., the number of laws passed per legislature), the complexity (e.g., the average length and readability of legislative texts), legislative coherence (e.g., the degree of interconnectedness between different laws and regulations), and legislative impact (e.g., the frequency and significance of legal changes). By tracking these metrics over time, stakeholders can assess legislative bodies' performance, identify improvement areas, and benchmark best practices in lawmaking.
 
We implemented these requirements to derive the Italian Legislative Property Graph (ILPG) in Neo4j, the most popular property graph database~\cite{PGranking,guia2017graph}. To achieve that, we first developed an end-to-end pipeline that regularly extracts textual information from the endpoint of the Italian legislation, i.e., Normattiva~\cite{normattiva}, through the laws published in AKN. We apply a set of transformations to achieve a consistent, interlinked representation in a property graph of the domain and we enrich it with some additional relevant information that can be easily accessed from official legislative endpoints. This resulted in the development of a complete and regularly updated property graph of the Italian national legislation, which is available at~\cite{dataGraph}.
Our contributions can be summarized as follows.
\begin{itemize}[leftmargin=4mm]
    \item We propose a property graph schema modeling legislative systems and their interconnections, whose generalization is backed by the adopted XML standard, Akoma Ntoso. 
    \item We implement the first end-to-end pipeline for building a Neo4j graph applying the aforementioned schema to a real-world use case, the Italian legislative system, also enriching it with country-specific features.
    \item We demonstrate the strength of our model by discussing relevant advanced queries that contribute to performing a comprehensive analysis of the Italian legislative systems.
\end{itemize}

\vspace{-1mm}

\section{Related Work}
The representation of legislative systems using graph databases has gained significant attention in recent years from both researchers and legislators. 
Most of these efforts have focused on using the Resource Description Framework (RDF) with several proposals put forward for utilizing RDF graphs to represent legislative systems~\cite{Junior2019UsingML}, with some preliminary prototypes being implemented in Italy~\cite{anelli2022navigating}, in Greece~\cite{angelidis2018a} and Spain~\cite{rodriguez2018spanish}. In such models, nodes, representing laws, are linked via RDF predicates representing relationships such as "\textit{amends}", "\textit{derives from}", "\textit{cites}", and so on. All these efforts represent a relevant step towards making governments more accountable to citizens;
they contribute to increasing transparency by facilitating the understanding of interconnections and allowing users to navigate the data in a user-friendly graphical form. However, each of such prototypes has limitations. 
Namely, the Spanish workflow~\cite{rodriguez2018spanish} uses named entity recognition techniques to build the RDF graph, potentially generating errors or omissions, given the complexity of correctly identifying laws without an identifier~\cite{detectionReferences,Sadeghian2018}. 
The Italian prototype~\cite{anelli2022navigating} focuses on developing tools for assisting the navigation of the legislative system. Still, the possibility of performing complexity analysis over the RDF-based legislative system is rather limited, as highlighted by their main use-case applications, which are more oriented to visualization~\cite{curtotti2012enhancing,curtotti2013software} and data access rather than focused on analyzing patterns and the complexity of the legislative system. 
In other works, parsers have been implemented to transform country-specific XML standards, representing legal knowledge, into RDF knowledge graphs, via the use of mapping languages~\cite{mappingXMLtoRDF}. Again, here is the idea of using such knowledge to support the development of a knowledge graph-based search system of the legal domain~\cite{crotti2020knowledge,Oliveira2023}.

Beyond visualization, previous works have also applied network analysis techniques to study citation patterns and dependencies within legislative systems, preliminarily investigating properties such as centrality, clustering, and community structure to uncover hidden relationships and dynamics~\cite{Bommarito2009,boulet2018network}. However, the absence of an underlying graph database capable of representing all properties and features of laws and citations forced such studies to simplify the data model, ignoring the complexity of the actual system which, for instance, distinguishes types of citations.

\vspace{-1mm}

\section{Legislative Property Graph Schema}
\label{sec:schema}

In this section, building on recent advancements in the adoption of common international XML-based law representation standards, we propose a legislative property graph schema that can combine the flexibility of its schema with the more natural expression of path-based queries in a property graph schema, offering the possibility of representing the complexity of legislative systems in a graph and in a format that allows an easy computation of systemic metrics and path traversal.

\vspace{-1mm}

\subsection{The XML Akoma Ntoso standard}
The adoption of an XML standard by an increased number of countries speeds up the process of building, analyzing, and comparing legislative systems. 
Akoma Ntoso is one of the most promising standards officially adopted throughout many countries and institutions.
Its main ability is to capture essential common features of law documents that are shared throughout different systems, such as the article-based splitting of laws or the modeling of references to other laws through standardized identifiers agreed on an international level~\footnote{In the EU, the European Legislation Identifier (ELI)~\cite{ELI}}. 
In general, AKN can offer advantages to various normative and regulatory documents by enabling a formal description of their framework, elements (such as attachments), and connections to and from other documents. 
The specifications of the AKN standard have also been approved by the OASIS body~\cite{aknAdoption}, recognizing that the standard meets the criteria of high quality and interoperability between legislative systems. 
This standard has also been adopted, even if not yet implemented, by the European Parliament~\cite{EUAKN}. This adoption will likely motivate other EU member states to standardize their systems toward this format. In the USA, efforts have been made to convert the code into AKN, promoted by the U.S. Library of Congress~\cite{AKNUSA}.
AKN has already been adopted and implemented in countries like Italy, through its official portal \textit{Normattiva}~\footnote{\url{https://www.normattiva.it/}}, the UK~\footnote{\url{https://www.legislation.gov.uk/}}, Switzerland~\footnote{Since May 30th, 2022, all new publications are in AKN \url{https://www.fedlex.admin.ch/eli}} and also by international institutions like the United Nations~\cite{AKNUN} and the FAO~\cite{palmirani2019akoma}.

\begin{table}
\centering
\resizebox{0.9\columnwidth}{!}{    
    \begin{tabular}{|c|c|l|}
    \hline
      &  \textbf{XML Tag}  &  \textbf{Content} \\
    \hline
        \multirow{6}{*}{\rotatebox[origin=c]{90}{\textbf{Metadata}}} & FRBRthis & Akoma Ntoso identifier for the act.\\
    & FRBRdate & Date of publication.\\
    & docType & Type of the act/law.\\
    & docTitle & Title of the law.\\
    & authorialNote & \makecell[l]{Text on additional relevant information about \\given aspects of the law.}\\
    \hline
    \hline
    \multirow{11}{*}{\rotatebox[origin=c]{90}{\textbf{Text Components}}} & preface / header &   \makecell[l]{Information related to the title of the document,\\ the identification numbers, the date of approval.}\\
    & preamble  & \makecell[l]{Introductory part of a document stating the legal\\ basis of a document.} \\
    & body  & \makecell[l]{Explicit presentation of hierarchy of parts, each \\identified with a name.} \\
    & article/section & \makecell[l]{Basic unit of the law body; it depends on the\\ specific legislative tradition.}\\
    & conclusions  & Block containing closing formulas and signature.\\
    & attachments & \makecell[l]{Documents that complete and integrate the\\ information of the main text.} \\
    \hline
    \hline
    \multirow{8}{*}{\rotatebox[origin=c]{90}{\textbf{References}}} & citations & \makecell[l]{References to the acts representing the legal basis.}\\
    & \makecell{activeMod} & \makecell[l]{Block for managing the modifications made by the\\ current document to another document.} \\
    & textualMod & Type of the modification to be applied.\\
    & source & \makecell[l]{Inside textualMod, it indicates the portion of the \\text where the modification is expressed.} \\
    & destination & \makecell[l]{Inside textualMod, it provides the document part \\ where the modification should be applied.}\\
    \hline
    \end{tabular}
    }
    \caption{Selected relevant (and shared) building blocks of Akoma Ntoso used to represent laws or acts produced by most legislation traditions.}
    \label{tab:aknstandard}
    \vspace{-8mm}
\end{table}

\smallskip
\noindent
\textbf{Main Building Blocks.} 
In Table~\ref{tab:aknstandard}, we report AKN building blocks that represent the basic structure of laws or acts that we will consider in the following sections to build the property graph.
The AKN schema is designed to capture different legislative traditions. For instance, in the \textit{preamble}, the formula describes the enacting sentences that are regular and fixed linguistic expressions for a specific tradition. 
The standard also provides a large number of tags for the parts that compose the \textit{body} (chapter, section, article, rule, etc) that denote the basic units of a legislative system. Without loss of generality, we adopt a unique tag, namely \textit{article}, to refer to the basic unit of a law. Note that, in the case of an Anglophone tradition, this would be more properly referred to as \textit{section} or \textit{rule} of the law.
Special attention is needed for the references. In fact, while the AKN standard dedicates specific blocks for active modifications and preamble citations, other citations might appear in the text inside a generic \textit{ref} tag. In such cases, we derive the source-destination pair by resorting to heuristics that detect which portion of the text is the source node and which is the destination node and avoid the generation of a duplicated reference that refers to amendments.
Finally, \textit{attachments} can be informative texts or technical data, for instance, tables, which, for practical reasons, do not appear in the \textit{body} of the law, or even other components such as an international agreement approved by the related act.

\subsection{Integration of additional data sources}
While comprehensive in many aspects, the AKN standard does not include certain information that may be useful to enhance the understanding of legislative frameworks. For instance, it overlooks key details regarding the broader legislative context within which laws are passed. Among these, in this work's pipeline, we mention 
the governing administration responsible for promoting a specific law or the legislature of reference, i.e., which parliament passed the laws. Country-specific sources need to be considered; in Section 4, we describe additional data sources considered for the Italian ILPG.

\subsection{Proposed Graph Schema}
\label{sec:graphschema}
\begin{figure}[t]
    \centering    \includegraphics[width=\linewidth]{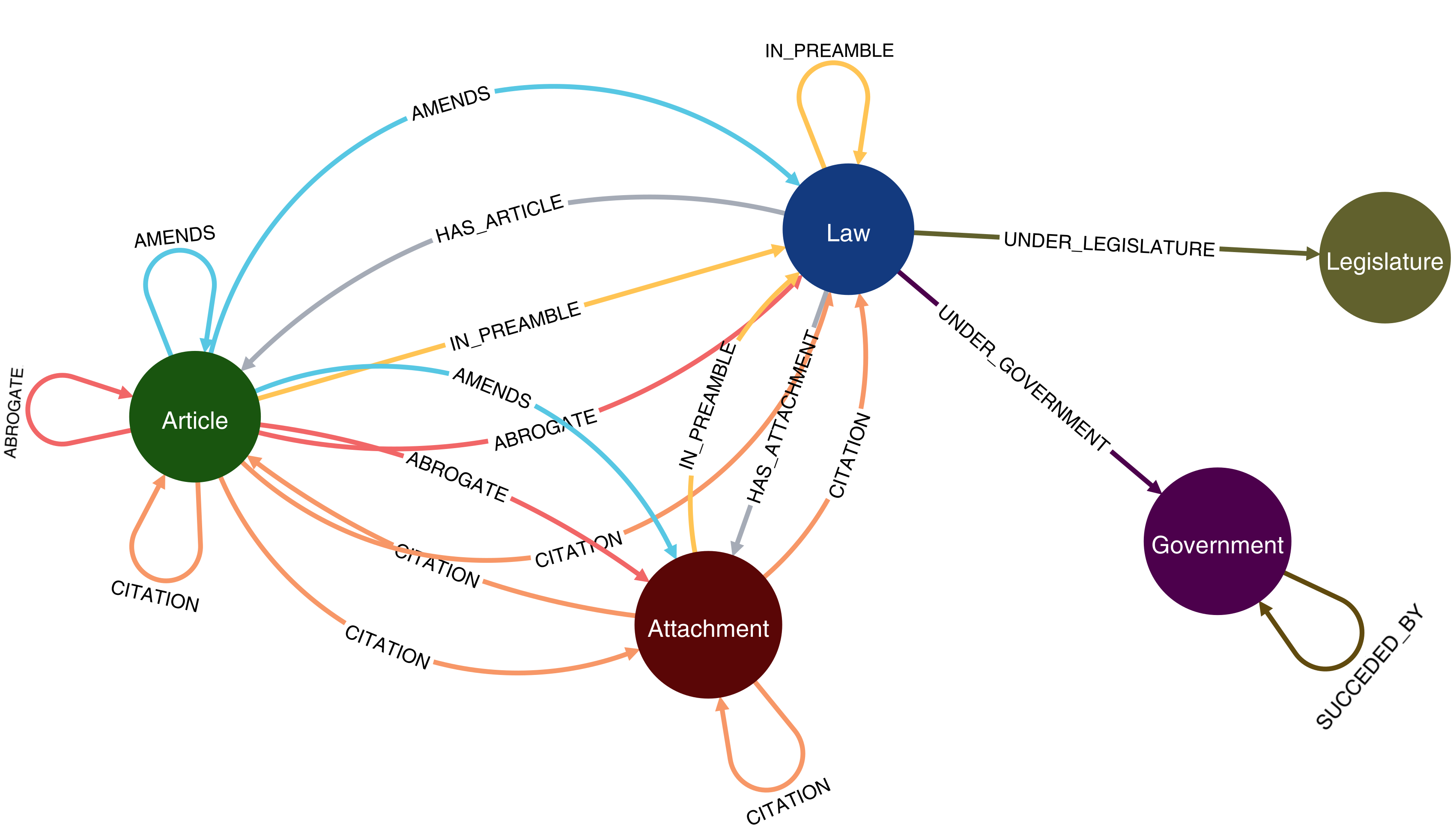}
    \includegraphics[width=\linewidth]{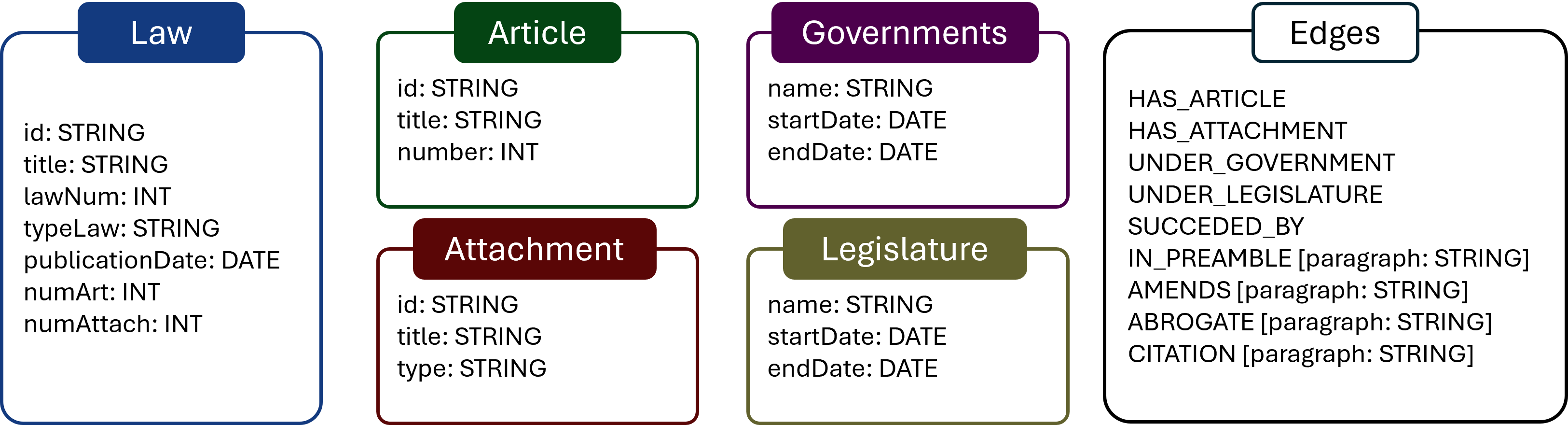}
    \caption{Graph schema visualization with the properties of nodes and edges; the related PG-Schema~\cite{angles2023pg} is provided on our repository~\cite{ourrepo}. Each node and edge can be enriched with additional properties, i.e., either country-specific features or other attributes that can be derived from the text.}
    \label{fig:graphschema}
    \vspace{-5mm}
\end{figure}

XMLs are semi-structured data models closely related to graphs and thus can be easily used to generate graphs~\cite{xmltoGraph}. Therefore, we will refer to XML tags to build the components of our property graph.
We consider Neo4j, the most used property graph database that stores data in a freely adjacent graph structure; nodes can be assigned multiple labels and attributes (called properties), and relationships can have a set direction and also include labels and attributes. 
Neo4j supports Cypher~\cite{cypher}, a declarative query language that allows expressive and efficient data querying in the property graph~\cite{sparqlcypher} and is very close to the soon-to-be standardized Graph Query Language (GQL)~\cite{gqlCypher}.
In Figure~\ref{fig:graphschema}, we depict our proposed baseline graph schema to model legislative acts by deriving the main components and attributes from the AKN standard and by including additional useful data about the legislative landscape, namely governments and legislatures under which laws are enacted. 

\smallskip \noindent \textbf{Schema Nodes}.
In our schema, we first model laws as nodes, assigning them metadata gathered from the corresponding tags, as described in Table~\ref{tab:aknstandard}. 
We generally call \textit{Article} the basic unit of law (i.e., article) and model it using dedicated nodes, directly connected to the law nodes through the \textit{HAS\_ART} relationship. We achieve this by leveraging the \textit{body} and \textit{article} XML tags. As articles are numbered progressively, we assign the corresponding identifier to each article node by joining the AKN identifier with the article number. Such a construct is, indeed, used to reference articles in other laws, thus allowing us to derive the edges.
Similarly, we model attachments as nodes linked through the \textit{HAS\_ATTACHMENT} edge. Their identifier is derived by joining the AKN law identifier with the numbered type of the attachment, which can be retrieved in the attachment tag. For instance, if the first attachment is a table, its identifier becomes: \textit{<AKN\_ID>\#Table 1}. 
Governments and legislatures are also modeled as additional nodes in the schema, linked to the laws according to their publication date, thus unlocking the possibility of connecting laws with the broader legislative context.

\smallskip \noindent \textbf{Reference Edges}.
We classify reference edges based on their role and nature. We first derive edges representing the legal basis of a law via the use of \textit{citation} tags. We represent them as \textit{IN\_PREAMBLE} directed edges, whose destination is the analyzed law and the source could be another law, an article, or even an attachment of another law. 
The edges might also have a property indicating the specific referenced \textit{paragraph} of the article. 
Then, we derive modification edges by leveraging the \textit{textualMod} blocks inside the \textit{activeMod} tag. Each of these tags represents a modification, i.e., an edge of our graph. The source nodes for these edges are articles of the analyzed law;
note that, as per normative drafting rules, attachments must contain content that can not be phrased in a normative way, thus excluding modification rules~\cite{draftnorms}.
The AKN standard already requires the type of modification to be indicated among a set of possible ones, namely substitution, insertion, split, join, renumbering, and repeal. We model such types by the \textit{AMENDS} and \textit{ABROGATE} edges, with the latter capturing repealing edges and the first encoding the remaining ones.
While preamble citations and modifications are captured by specific tags in the standard, other references can be found throughout the text. 
This is the case of more generic citations of other laws, articles, attachments, or paragraphs required to complement the text. We model them in our schema as \textit{CITATION} edges.
For instance, a reference is required when providing definitions of terms used in the text and that exemplify a certain category, e.g., what an industrial consortia is.
Still, the AKN standard helps us detect additional references as each citation is captured inside a \textit{ref} container, including the AKN identifier. Therefore, such edges can be gathered by parsing the text of the law: any citation of this category would have either an article or an attachment of the law as its source.  

\section{The Italian Legislative Property Graph (ILPG)}


Here, we apply the graph schema presented in Section~\ref{sec:schema} to a real-world use case, the Italian legislative system.
The modern Italian legislative system can be dated back to the adoption of the republican Constitution in 1948; this cutoff date can be used to discard obsolete laws referring to the Kingdom period.
All laws are publicly available through the \textit{Normattiva} portal, which already implements the Akoma Ntoso standard. To build the property graph, we gathered all the laws published after the Constitution and leveraged the XML standard to derive the set of nodes and edges, as proposed in Section~\ref{sec:graphschema}. 

\subsection{ETL Pipeline}
\label{sec:etlpipeline}
\begin{figure*}
    \centering
    \includegraphics[width=0.9\linewidth]{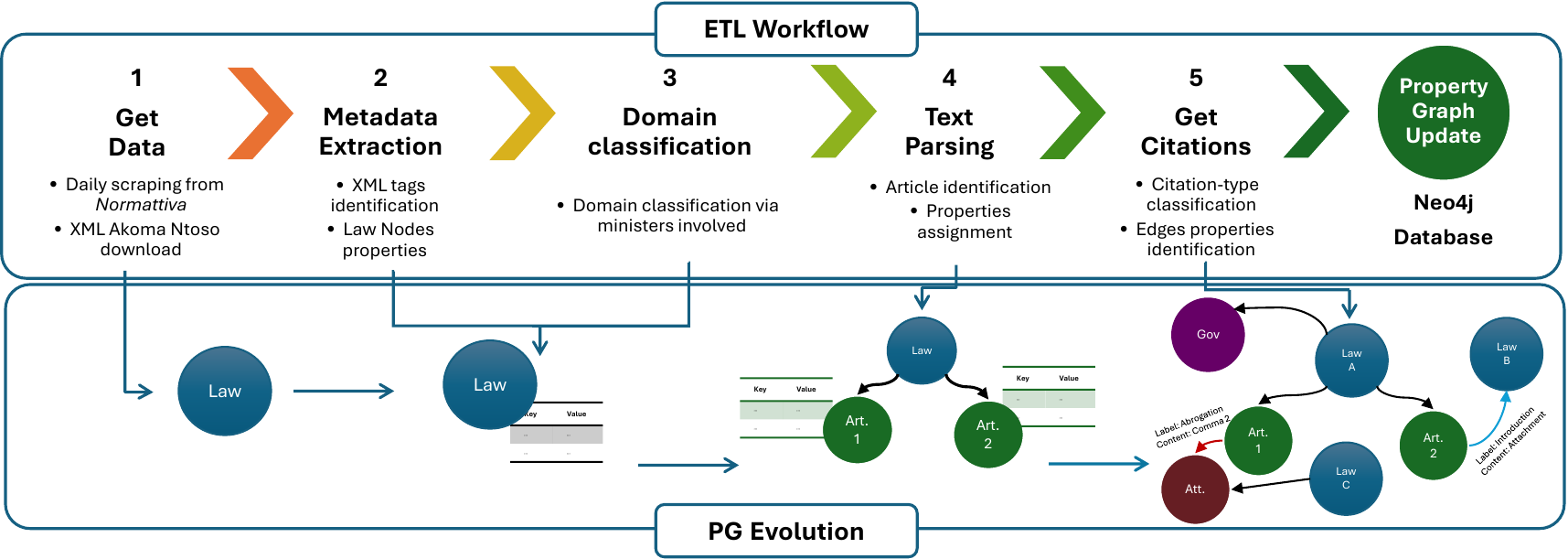}
    \caption{ETL pipeline to build the Italian Legislative Property Graph. At each step, new components of the PG are added.}
    \label{fig:etl}
    \vspace{-3mm}
\end{figure*}

The data provision step is represented in Figure~\ref{fig:etl}; it follows the ETL paradigm to build and update a legislative property graph. The pipeline runs on a daily basis and updates the property graph database on Neo4j. The graph is available at~\cite{dataGraph}.

\smallskip \noindent \textbf{1. Get Data}. The first step of the ETL pipeline is to ingest the data, i.e., the laws. Italian laws are available on the Normattiva portal, which is updated as soon as new laws are enacted. Our graph is updated automatically by downloading -daily- newly enacted laws. As in most legislation systems, we can distinguish between versions of the law. A law can evolve; for instance, it can be amended or partially repealed, leading to multiple versions of the same law, the so-called \textit{in-force} laws, according to a specific date. In our system, we track the evolution of laws through edges. To this end, we need the original version of acts (their first version after publication), as any change will be captured by ex-post incoming edges. 

\smallskip \noindent \textbf{2. Metadata Extraction}. For each law, relevant metadata are extracted by leveraging the corresponding AKN tags. Such information is assigned to laws as node attributes. More specifically, we retrieve the title, the date of publication, the type of act, and the AKN identifier from the corresponding tags (described in Table~\ref{tab:aknstandard}). 
In addition, we distinguish between the publication date and the in-force date: each law enters into force after a certain period since its publication on the official gazette. Such information can be derived from an \textit{authorialNote} tag in the \textit{preamble}, denoted as `\textit{Entrata in vigore del provvedimento}', i.e., in-force date of the law. Both features can be added as novel attributes to the law nodes. Finally, we also add the number of articles and attachments by counting the corresponding number of tags in the text.

\smallskip \noindent \textbf{3. Domain Classification}. Next, our pipeline enriches each law with its \textit{domain}. With domains, we refer to the government departments involved in the law, which can be useful for gaining insights about the domain of legislative activities, as we shall see in the next section.
The domains can be derived by analyzing the \textit{conclusions} of the law, where involved ministries sign the law. By parsing such text, we gather the ministries' surnames. This is done by leveraging the list of ministries in charge at the time of the law's publication date and checking which surname is present in the conclusions. 

In the case of the Italian legislation, such additional data can be obtained by querying the data provided by the lower house of the parliament, i.e., Camera dei Deputati, which makes related datasets available. Additionally, as the department's name might slightly change over time, we group and categorize them via a keyword-based fuzzy matching dictionary, which maps into categories the words that may appear in the department's name. For instance, `treasury', `finance', and `economics' have been associated with different departments at different times: here, these names are all mapped to the `economic' class.


\smallskip \noindent \textbf{4. Text Parsing}. At this step of the pipeline, we have created the law nodes enriched with useful attributes. The AKN standard allows us to go deeper into the granularity of information,
it defines the basic units of a law. In the Italian system, each law can be decomposed into \textit{articles}, each of which treats a different aspect of the law, also captured by an \textit{epigraph}, i.e., the title of the article. 
As articles are numbered, each of them can be identified by the law ID followed by its number.
A similar parsing can be applied to attachments, which we model into a different node type with specific features. While articles can only be textual, the form of attachment might vary from text to tables and schema, integrating the law. 

\smallskip \noindent \textbf{5. Get Citations}. The last step of our ETL pipeline retrieves the citations that connect laws, potentially via articles. As described in Section~\ref{sec:graphschema}, tags are leveraged to obtain the references and to classify them into the corresponding type. The only exceptions to the schema are the \textit{INTRODUCE} edges, which are a special type of insertion in the Italian legislation that introduce additional integrative articles in the law and are denoted by the number of the in-force article integrated with a Latin numeral progressive adverb.

\subsection{Main Features of the ILPG}
Besides generic common features, the Italian legislative system, like any other system, has its unique characteristics. First, laws follow a progressive year-based numbering, meaning that each law is assigned a number depending on the publication date and resetting each year.
Some laws are classified as constitutional laws, i.e., the ones that amend the Constitution. This distinction is required since other laws are assigned year-based progressive numeric identifiers, while constitutional ones follow a dedicated progressive numbering, thus affecting queries that leverage the law number as a filtering parameter. 

\begin{table}[t]
\small
    \centering
    \begin{tabular}{|l|c|c|c|}
    \hline
       \multicolumn{1}{|c}{\textbf{Graph Metric}} & \textbf{Laws}  & \textbf{Articles} & \textbf{Attachments} \\
    \hline
        Total number of nodes & 74020 & 317563 & 20064 \\
        In-degree (Preamble Edges) & 0.75 & 0.13 & 0.01\\
        In-degree (Citation Edges) & 2.40 & 0.58 & 0.80 \\
        In-degree (Amendment Edges) & 0.09 & 0.22 & 0.35\\
        In-degree (Abrogation Edges) & 0.007 & 0.19 & 0.13\\
        In-degree (Introduction Edges) & - & 0.016 & 0.025\\
    \hline
    \end{tabular}
    \caption{Relevant statistics characterizing the nodes and edges of the Italian Legislative Property Graph.}
    \label{fig:IKGdimensions}
    \vspace{-8mm}
\end{table}

In Table~\ref{fig:IKGdimensions}, we illustrate the Italian Legislative Property Graph dimensions. In total, we modelled over 400k nodes and over 1 million edges, including both references and hierarchy edges. Regarding references, we report the total in-degree value for each reference edge category, i.e., the number of times a node is a source for the preamble, citation, amendment, abrogation, and introduction edges. Such values tell us how laws, articles, and attachments are commonly used in the legislative system. For instance, attachments are rarely used as the legal basis for new laws, whereas in most cases, we can find entire laws.

\smallskip \noindent \textbf{Temporal Features}. A major shift in law enactment can be observed in the mid-80s, with a change regarding how laws are drafted. Laws started to become fewer but longer, with more articles and attachments per law. While analyzing the motivations behind such change is beyond the scope of our paper, we note that such behavior must be considered when performing queries, whose results might be strongly influenced by this feature. For instance, a simple query that finds the governments with the most produced laws might be biased.
We also note that, with the exception of a couple of relevant laws, namely, the Civil and Penal Codes, we excluded laws enacted before the Italian constitution.
Thus, edges whose destination nodes are excluded are not represented in our database; this can potentially affect queries leveraging citation edges.
Finally, special attention should be given to so-called \textit{Decreti Semplificazione}, i.e., lump sum decrees enacted to abrogate outdated laws. Therefore, such decrees contain many repealing rules that aim to ``clean'' the legislative landscape from obsolete acts: whenever performing queries involving abrogate edges, users might want to filter out such laws, namely laws 2008/112, 2010/66 and 2010/212.

\begin{figure*}[t]
    \centering
    \subfloat[Published Laws (\ref{q:lawProduction})]{\includegraphics[width =0.24\linewidth]{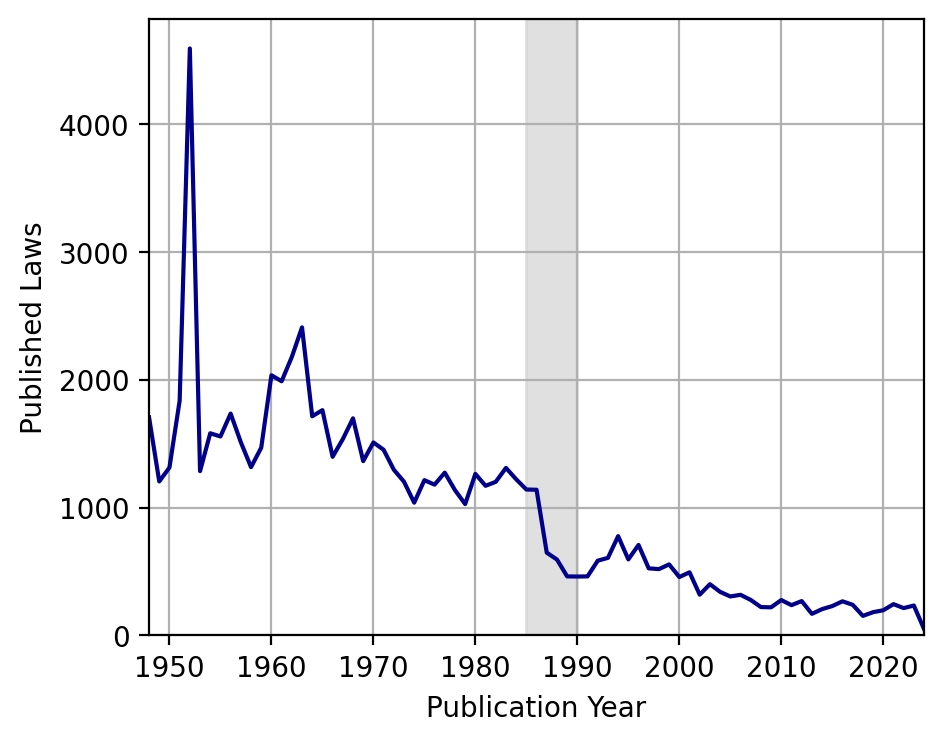}\label{fig:query1}}
    \subfloat[Never Cited Laws (\ref{q:neverCited})]{\includegraphics[width =0.24\linewidth]{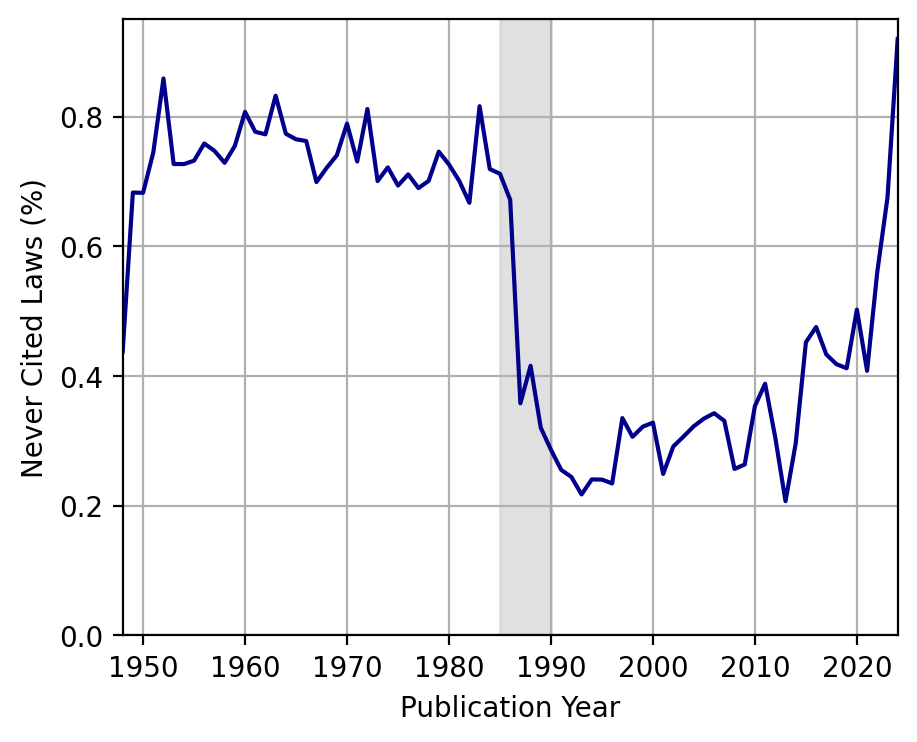}\label{fig:query2}}
    \subfloat[Outdated Laws (\ref{q:outdatedLaws})]{\includegraphics[width =0.24\linewidth]{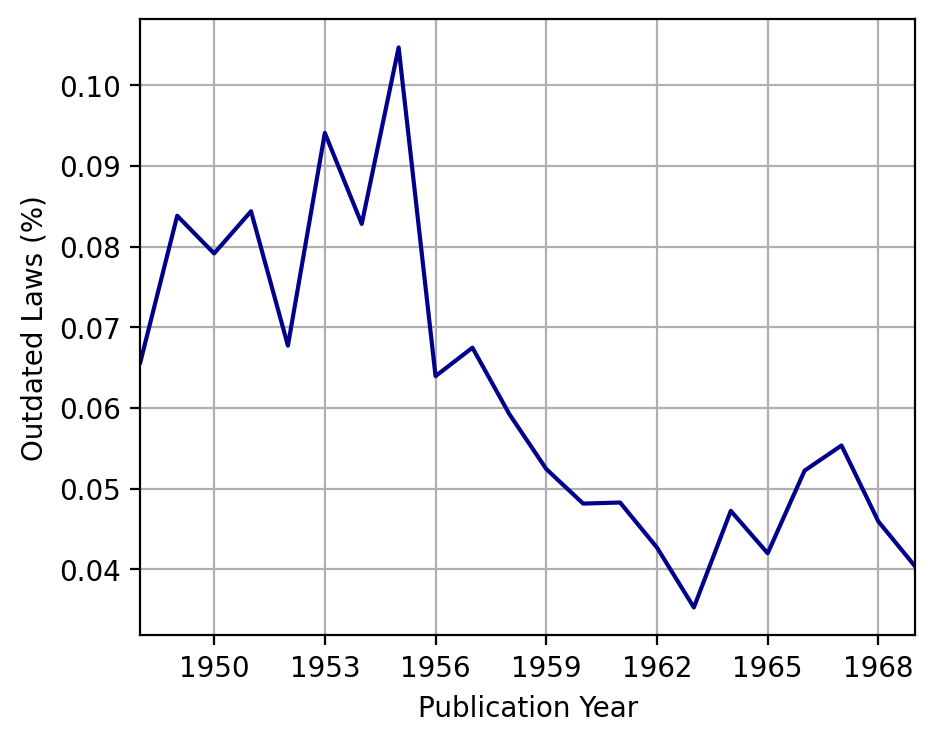}\label{fig:query3}}
    \subfloat[Stock in-force Laws (\ref{q:stockLaws})]{\includegraphics[width =0.24\linewidth]{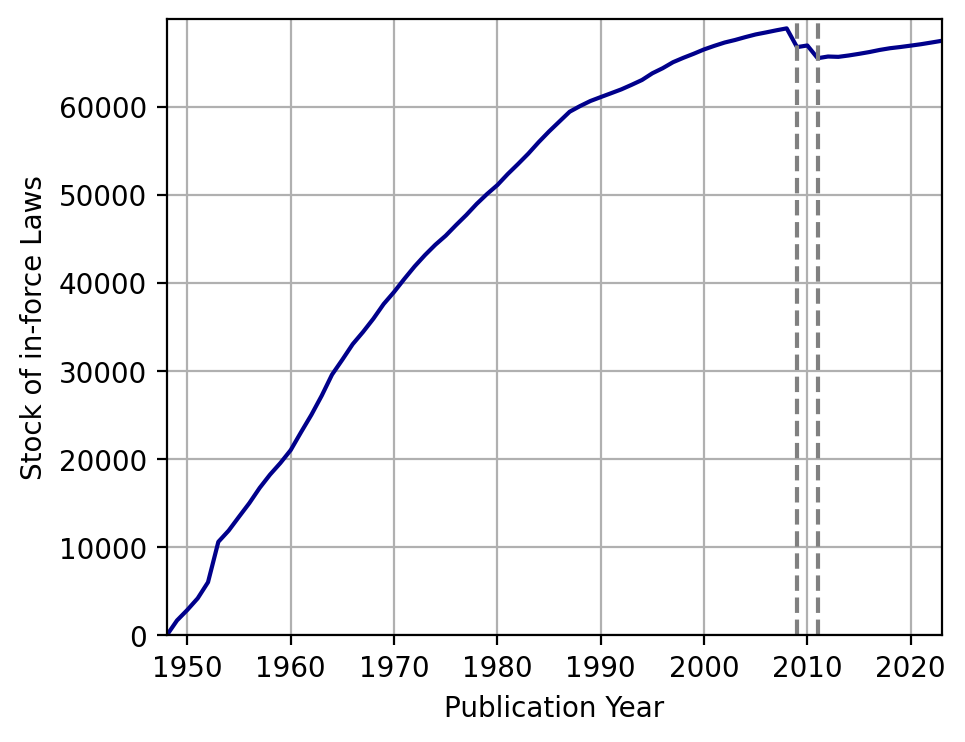}\label{fig:query4}}
    \caption{Panel (a) shows published laws resulting from \ref{q:lawProduction}, the initial peak is due to the shift from the monarchy to the republic. Panel (b) shows the the result of \ref{q:neverCited} (the fraction of laws published at a given time and never cited). Both
    \ref{q:lawProduction} and \ref{q:neverCited} display a change in drafting laws occurred around the '80s, with \ref{fig:query2} also showing that many recent laws are not yet cited.
    In panel (c), we find the result of \ref{q:outdatedLaws}; 
    here, we considered all laws published before the cut-off date 1970, and then we used D=1992 (a significant year in Italian politics -- the start of the so-called second republic) in order to compute the fraction of laws not cited after D.  
    Panel (d) responds to  \ref{q:stockLaws}; here we highlighted the drops corresponding to simplification decrees of 2008 and 2010.}
    \label{fig:queryresults}
\end{figure*}

\section{Queries and Graph Patterns}
\label{sec:queries}
By leveraging the ILPG, we explore and demonstrate the analytical power unlocked by modeling legislative systems within our property graph framework. To this aim, we examine some types of structural queries facilitated by this approach.
We provide examples of queries to illustrate the possibility of identifying complex patterns that allow researchers/stakeholders to gain a deeper understanding of the represented legislative landscape; related Cypher scripts are available in our repository \cite{ourrepo}.

\subsection{Law-centric Queries}
A set of queries can be designed to derive general insights into the legislative system, by filtering and aggregating attributes (for instance by year, legislature, or government). 
We illustrate the derived plotted results in Figure~\ref{fig:queryresults}.

\begin{itemize}[leftmargin=5mm]
    \addtocounter{querycount}{1} \myitem{Q\thequerycount} \label{q:lawProduction} \textit{Laws enacted per year}, i.e., a count over the publication year. 
    \addtocounter{querycount}{1} \myitem{Q\thequerycount} \label{q:neverCited} \textit{Laws never cited after publication}, i.e., laws that have not been referenced in any preamble or have not received any amendment, abrogation, introduction or other citations.
    \addtocounter{querycount}{1} \myitem{Q\thequerycount} \label{q:outdatedLaws} \textit{Detection of outdated laws},  i .e., laws that have ceased to be cited after a given time. To detect them, we first derive laws that have been published before a cut-off date. 
    Then, by defining another date D subsequent to the cut-off date, e.g., a significant event in politics, we extract the set of laws that have been cited by any law published after D, consequently also detecting the set of laws that has not been cited after D.
    \addtocounter{querycount}{1} \myitem{Q\thequerycount} \label{q:stockLaws} \textit{Stock of in-force laws}. As a measure of the complexity of the observed legislative system, we track the total number of laws in force on a certain date.
    This requires detecting which laws have not been abrogated until that moment. Note that, in the case of Italian law, the official data source Normattiva allows one to select a specific point in time to view whether a law is in force or repealed. However, all laws must be gathered again with the desired date selected. Instead, by leveraging \textit{abrogate} edges, in the ILPG we can derive which laws have been repealed, i.e., whenever all its articles have been abrogated, 
    or whenever there is a direct abrogation of the entire law.
\end{itemize}

\subsection{Legislative Pattern Queries}
Here, we describe a set of queries that leverage the additional data gathered on top of laws, demonstrating the achieved interoperability that unlocks the derivation of insights about the broader legislative system.

\begin{itemize}[leftmargin=5mm]
    \addtocounter{querycount}{1} \myitem{Q\thequerycount} \label{q:errorCitation} \textit{Articles that have been used as legal basis when already abrogated}. By tracking labeled edges in the property graph, we detect if and which articles have been cited after their abrogation, detecting errors in the legislative system. Through this query, we detected 144 citation errors, quite uniformly distributed across the years. 
    \addtocounter{querycount}{1} \myitem{Q\thequerycount} \label{q:tableAttach} \textit{Number of tabular attachments per law for each domain}. By leveraging the derived \texttt{domain} attribute, we can analyze features characterizing laws over different topics, for instance, by computing the frequency of tabular attachments. The result of such a query illustrates how table attachments are very frequent (372) whenever the `Home Office', `education', and `economics' departments are jointly involved, more than three times w.r.t. strictly economic laws (117).
    \addtocounter{querycount}{1} \myitem{Q\thequerycount} \label{q:decreeConversion} \textit{Average number of amendments to government decrees}. It measures the tendency of the parliament to apply amendments and/or abrogations to so-called decrees, which, in the Italian system, are directly enacted by the government for emergency needs and have to be converted into laws in 45 days. 
    The number of amendments and abrogations can be computed by counting the incoming edges from the conversion law node to the decree node. First, conversion laws have to be detected by leveraging the aforementioned pattern. Then, for each government, we can count the average number of amends and abrogate edges that are directed towards the government decree. Table~\ref{tab:results_query2} (left) shows the top three results.
\end{itemize}

\begin{figure}[t]
    \centering
    \includegraphics[width = 0.85\linewidth]{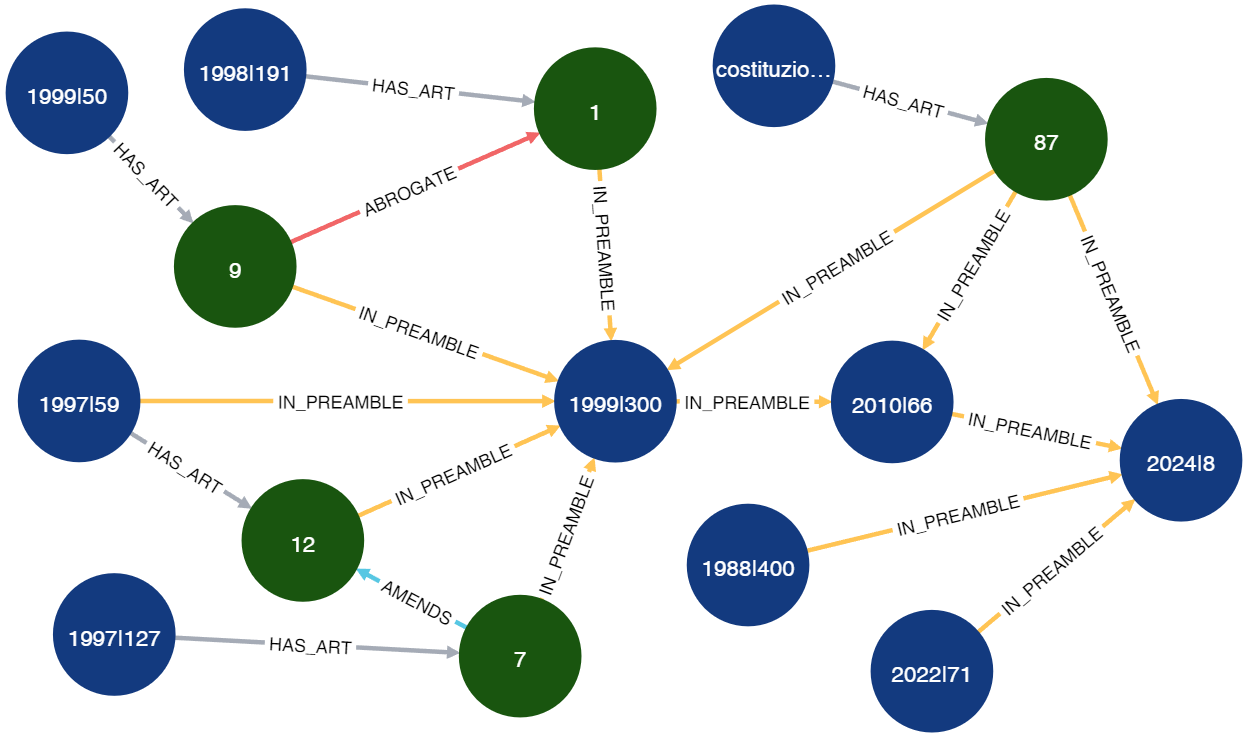}
    \caption{Direct and indirect legal basis of the law of interest 2024/8 (blue node on the right-end). Its preamble cites three laws (2010/66, 1988/400, 2022/71) and an article of the Italian Constitution (87). Law 2010/300 has its foundation in law 1999/300, which, in turn, is based on articles of other laws (1 of 1998/191, 9 of 1999/50, 12 of 1997/59, and 7 of 1997/127). Thus, the latter laws are also indirectly relevant to law 2024/8.}
    \label{fig:dependencies}
\end{figure}

\begin{table}
\small
    \subfloat{\begin{tabular}{|c|c|}
    \hline
      \textbf{Government} & \textbf{AVG(N)} \\
    \hline
       Rumor-II  & 20 \\
    \hline
       Moro-III  & 15.2 \\
    \hline
       Berlusconi-IV  & 15.16 \\
    \hline
       \ldots & \ldots \\
    \hline
    \end{tabular}}
\quad
    \subfloat{\begin{tabular}{|c|c|c|}
    \hline
      \textbf{SourceGov} & \textbf{DestGov} & \textbf{A} \\
    \hline
       Letta-I  & Berlusconi-IV & 0.7 \\
    \hline
       Berlusconi-III  &  Berlusconi-II & 0.68 \\
    \hline
       Monti-I  & Berlusconi-IV & 0.41 \\
    \hline
       \ldots & \ldots & \ldots \\
    \hline
    \end{tabular}}
    \caption{Top results for queries \ref{q:decreeConversion} (left table) and \ref{q:repealingAttitude} (right table). $N$ denotes the number of modifications that are applied to government decrees. $A$ denotes a government's number of abrogations - per day in charge - made to one of its two adjacent predecessors.}
    \label{tab:results_query2}
\vspace{-4mm}
\end{table}

\vspace{-1mm}

\subsection{Legislative Chain Paths}
Our last queries fully leverage the property graph model and the efficiency in path traversal. 
\begin{itemize}[leftmargin=5mm]
    \addtocounter{querycount}{1} \myitem{Q\thequerycount} \label{q:legalbasis} \textit{Dependencies and legal basis of a given law}. By traversing paths of unpredictable length through chains of laws, we can derive the sub-graph of dependencies for a law, offering a representation of all its legal basis, enabling further analysis on the resulting nodes, as in Figure~\ref{fig:dependencies}.
    \addtocounter{querycount}{1} \myitem{Q\thequerycount} \label{q:repealingAttitude} \textit{Analysis of government repealing attitude toward its predecessors}. The tendency of governments to amend or repeal laws promoted by previous governments can be easily measured by leveraging the chain of adjacent governments to detect the edges of interest. Such metrics can also be presented in relative terms, for instance, by leveraging the duration of a government. The results of such a query show a government's attitude to change and/or repeal acts and rules enacted by their direct predecessors, for instance, potentially detecting continuity behaviors beside different political colors.
    \addtocounter{querycount}{1} \myitem{Q\thequerycount} \label{q:shortestPath} \textit{Degrees of separation from the Constitution}.  By computing the shortest path between two nodes, we can derive, for each law, its separation degree from the Constitution (the fundamental law of the Italian Republic). 
    Typically, laws that are far from the Constitution are less important than others.
    We empirically verified this claim by deriving the average degree of distance of all laws from the Constitution w.r.t. the type of law. We indeed found that \textit{ordinances} -- which are merely administrative acts -- are on average more distant (3.2 degrees) than \textit{government decrees} (2 degrees) (which are the closest ones).
\end{itemize}
Additional queries can also be used to derive sub-graphs 
on which network analysis measures can be computed, e.g., for detecting the most important laws in terms of legal basis through centrality algorithms.


\vspace{-2mm}

\subsection{Discussion and Limitations}
By exemplifying some relevant queries, we demonstrated how modeling the legislative system into a property graph improves and supports the analysis of the legislative landscape. 
The adoption of a graph database preserves the possibility to extract and explore most of the relevant information that a relational database can offer;
this is the case of computing aggregations to explore the dimensions describing tendencies and features of the legislative landscape---see \ref{q:lawProduction}, \ref{q:tableAttach}, or \ref{q:legalbasis}, where results are accompanied by useful visualization (Figure~\ref{fig:dependencies}). 
In addition, we illustrated how our data model helps derive useful insights and metrics by handling non-trivial aspects of the law (\ref{q:neverCited}, \ref{q:stockLaws}, 
\ref{q:repealingAttitude}, \ref{q:shortestPath}) and, especially, in the detection of errors (\ref{q:errorCitation}), inefficiencies (\ref{q:outdatedLaws}, \ref{q:decreeConversion}) that could help in simplifying the overall system complexity as well as in the drafting of new laws. Such queries are enabled by the property graph structure, which supports an efficient graph-path traversal and shows high flexibility in nodes and especially edges properties.
The proposed queries exemplify analysis scenarios over a real-world legislative property graph, while we leave policy-relevant discussions over the findings of some proposed queries to domain experts. 
For instance, some laws detected by \ref{q:outdatedLaws} might still be relevant, although they are not used as legal basis anymore.

Finally, we mention that our ETL pipeline to build and update the ILPG is based on an XML standard that has been adopted and is produced by the Italian bureaucracy. Although we implemented automatic quality checks on the input XML data, they remain subject to errors, especially for older laws. For instance, in some of these laws, the \textit{preamble} or the \textit{conclusion} tags were missing and we had to resort to heuristics to identify and split those building blocks.


\section{Conclusion}



In this paper, we presented a comprehensive property graph schema designed to represent legislative systems and the interconnections between laws, the first model to reach a deeper granularity, by also capturing basic law units and integrating other data sources. To the best of our knowledge, our schema is the first to be built on top of the recently adopted XML machine-readable standard for representing law documents, Akoma Ntoso. We implemented the schema over the Italian legislation, and we built an ETL pipeline that extracts law information and models them in a Neo4j database, the most popular property graph solution. Given the international adoption of such a standard, we think the same pipeline can be easily applicable in other systems, with few adjustments that account for country-specific attributes; as a consequence, information retrieved from different legislative graphs will be potentially comparable for extracting further insights. 
Finally, we explored how this model allowed us to run non-trivial queries, enabling the detection of significant patterns and insights within the legislative corpus. For instance, our queries facilitated the identification of temporal trends, features and quantitative metrics which are notably hard to express in other data models, as they rely on graph patterns and traversals.

\smallskip \noindent
\textbf{Resources.} The Cypher scripts with the presented queries and a PG-Schema~\cite{angles2023pg} formalization for our graph schema are available in our repository \cite{ourrepo}. The graph is publicly available at~\cite{dataGraph}.

\smallskip \noindent
\textbf{Acknowledgements.}
S.C. is supported by the PNRR-PE-AI FAIR project, funded by the NextGenerationEU program. A.C. kindly acknowledges INPS for the funding of his Ph.D. program. We thank Claudio Michelacci and Luigi Guiso from EIEF for the fruitful discussions and for indicating the primary data source for the ILPG.

\clearpage

\bibliographystyle{ACM-Reference-Format}
\bibliography{biblio}

\end{document}